\documentclass[12pt,preprint]{aastex}
\usepackage{psfig}
\usepackage{epsfig}
\usepackage{color}
\usepackage{amsmath,amssymb}
\usepackage{graphicx}

\def\arcs{\ifmmode ^{\prime\prime} \else $^{\prime\prime}$}
\def\simlt{\lower.5ex\hbox{$\; \buildrel < \over \sim \;$}}
\def\simgt{\lower.5ex\hbox{$\; \buildrel > \over \sim \;$}}
\def\hst{{\em HST}}
\def\hnot{\ifmmode H_0 \else H$_0$\fi}
\def\msun{\ifmmode {M_\odot} \else $M_\odot$\fi}
\def\lsun{\ifmmode {L_\odot} \else $L_\odot$\fi}
\def\kms{km s$^{-1}$}
\def\msunyr{\ifmmode {\rm M_\odot~yr^{-1}}\else${\rm M_\odot~yr^{-1}}$\fi}
\def\lam{\ifmmode {\lambda} \else {$\lambda$} \fi}
\def\lamLlam{\ifmmode \lambda L_{\lambda}(5100) \else 
	 {$\lambda L_{\lambda}(5100)$} \fi}
\def\nuLnu{\ifmmode \nu L_{\nu}(5100) \else {$\nu L_{\nu}(5100)$} \fi}

\def\rblr{\ifmmode {R_\mathrm{BLR}} \else $R_\mathrm{BLR}$ \fi}
\def\mdoto{\ifmmode {\dot{M}_0} \else $\dot{M}_0$ \fi}
\def\ilam{\ifmmode {I_\lambda} \else $I_\lambda$ \fi}
\def\flam{\ifmmode {F_\lambda} \else $F_\lambda$ \fi}
\def\inu{\ifmmode {I_\nu} \else $I_\nu$ \fi}
\def\fnu{\ifmmode {F_\nu} \else $F_\nu$ \fi}
\def\yr{\ifmmode {\rm yr} \else yr \fi}
\def\cm{\ifmmode {\rm cm} \else cm \fi}
\def\cmmitwo{\ifmmode \rm cm^{-2} \else $\rm cm^{-2}$\fi}
\def\cmmithree{\ifmmode \rm cm^{-3} \else $\rm cm^{-3}$\fi}
\def\cmps{\ifmmode \rm cm~s^{-1}\else $\rm cm~s^{-1}$\fi}
\def\cmpsps{\ifmmode \rm cm~s^{-2}\else $\rm cm~s^{-2}$\fi}
\def\kms{\ifmmode \rm km~s^{-1}\else $\rm km~s^{-1}$\fi}
\def\kmspmpc{\ifmmode \rm km~s^{-1}~Mpc^{-1} \else
  $\rm km~s^{-1}~Mpc^{-1}$\fi}
\def\ergps{\ifmmode \rm erg~s^{-1} \else $\rm erg~s^{-1}$ \fi}
\def\ergpspcm{\ifmmode \rm erg~s^{-1}~cm^{-2} \else $\rm erg~s^{-1}~cm^{-2}$ \fi}
\def\ergpspcmphz{\ifmmode \rm erg~s^{-1}~cm^{-2}~Hz^{-1} \else $\rm
erg~s^{-1}~cm^{-2}~Hz^{-1}$ \fi}
\def\ergpspcmpa{\ifmmode \rm erg~s^{-1}~cm^{-2}~\AA^{-1} \else $\rm
erg~s^{-1}~cm^{-2}~\AA^{-1}$ \fi}
\def\ergpsphz{\ifmmode \rm erg s^{-1} Hz^{-1} \else
  $\rm erg s^{-1} Hz^{-1}$ \fi}
\def\mbh{\ifmmode M_{\bullet} \else $M_{\bullet}$\fi}
\def\msigma{\ifmmode M_{\sigma} \else $M_{\sigma}$\fi}
\def\dmsigma{\ifmmode \delta_{M\sigma} \else $\delta_{M\sigma}$\fi}
\def\mbulge{\ifmmode M_{\mathrm{bulge}} \else $M_{\mathrm{bulge}}$\fi}
\def\mgal{\ifmmode M_{\mathrm{gal}} \else $M_{\mathrm{gal}}$\fi}
\def\lgal{\ifmmode L_{\mathrm{gal}} \else $L_{\mathrm{gal}}$\fi}
\def\lbulge{\ifmmode L_{\mathrm{bulge}} \else $L_{\mathrm{bulge}}$\fi}
\def\mgalstar{\ifmmode M^*_{\mathrm{gal}} \else $M^*_{\mathrm{gal}}$\fi}
\def\mbhstar{\ifmmode M^*_{\bullet} \else $M^*_{\bullet}$\fi}

\def\mbhsigstar{\ifmmode M_{\bullet} - \sigma_* \else $M_{\bullet} - \sigma_*$ \fi}
\def\lsigstar{\ifmmode L_{\mathrm{bulge}} - \sigma_* \else $L_{\mathrm{bulge}} - \sigma_*$ \fi}
\def\mbhl{\ifmmode M_{\bullet} - L_\mathrm{bulge} \else $M_{\bullet} - L_\mathrm{bulge} $ \fi}
\def\deltalogmbh{\ifmmode \Delta~{\mathrm{log}}~M_{\bullet} \else 
  $\Delta$~log~$M_{\bullet}$\fi}

\def\sigstar{\ifmmode \sigma_* \else $\sigma_*$\fi}
\def\sigthree{\ifmmode \sigma_{\mathrm{[O~III]}} \else $\sigma_{\mathrm{[O~III]}}$\fi}
\def\sigtwo{\ifmmode \sigma_{\mathrm{[O~II]}} \else $\sigma_{\mathrm{[O~II]}}$\fi}
\def\signl{\ifmmode \sigma_{\mathrm{NL}} \else $\sigma_{\mathrm{NL}}$\fi}
\def\wthree{\ifmmode {\rm FWHM({[O~III]})} \else $FWHM({[O~III]})$ \fi}
\def\wtwo{\ifmmode {\rm FWHM({[O~II]})} \else $FWHM({[O~II]})$ \fi}
\def\mthree{\ifmmode M_{\mathrm [O~III]} \else $M_{\mathrm [O~III]}$ \fi}
\def\mtwo{\ifmmode M_{\mathrm [O II]} \else $M_{\mathrm [O II]}$ \fi}

\def\lbreak{\ifmmode L_{\mathrm{break}} \else $L_{\mathrm{break}}$\fi}
\def\lcut{\ifmmode L_{\mathrm{cut}} \else $L_{\mathrm{cut}}$\fi}
\def\led{\ifmmode L_{\mathrm{Ed}} \else $L_{\mathrm{Ed}}$\fi}
\def\lbol{\ifmmode L_{\mathrm{bol}} \else $L_{\mathrm{bol}}$\fi}

\def\hbeta{\ifmmode {\rm H}\beta \else H$\beta$\fi}
\def\mgii{\ifmmode {\rm Mg{\sc ii}} \else Mg~{\sc ii}\fi}

\newcommand{\gae}{\lower 2pt \hbox{$\, \buildrel {\scriptstyle >}\over {\scriptstyle \sim}\,$}}
\newcommand{\lae}{\lower 2pt \hbox{$\, \buildrel {\scriptstyle <}\over {\scriptstyle \sim}\,$}}
\newcommand{\oiii}{{\sc [O~iii]}}
\newcommand{\oii}{{\sc [O~ii]}}

\slugcomment{Submitted to The Astrophysical Journal}

\shorttitle{The Largest Velocity Dispersion Galaxies}
\shortauthors{Salviander et al.}

\begin{document}

\title{In Search of the Largest Velocity Dispersion Galaxies}

\author{S. Salviander, G.~A. Shields, K. Gebhardt}
\affil{Department of Astronomy, University of Texas, Austin, TX 78712}

\and

\author{M. Bernardi, J.~B. Hyde}
\affil{Department of Physics and Astronomy, University of Pennsylvania, Philadelphia, PA 19104}

\begin{abstract}

We present Hobby-Eberly Telescope (HET) observations for galaxies at redshift $z < 0.3$ from the Sloan Digital Sky Survey (SDSS) showing large velocity dispersions while appearing to be single galaxies in \hst\ images. The high signal-to-noise HET spectra provide more definitive velocity dispersions. The maximum velocity dispersion we find is $\sigstar = 444~\kms$. Emission-line widths in QSOs indicate that black holes can exist with masses \mbh\ exceeding 5~billion \msun, implying $\sigstar > 500~\kms$ by the local \mbhsigstar\ relationship. This suggests either that QSO black hole masses are overestimated or that the black hole - bulge relationship changes at high black hole mass. The latter option is consistent with evidence that the increase in \sigstar\ with luminosity levels off for the brightest elliptical galaxies.

\end{abstract}


\section{Introduction}
\label{s:intro}

The use of broad emission lines in QSOs to measure black hole mass, \mbh, allows study of black hole demographics over a large volume of space. The largest black hole masses range up to $10^{10}$ \msun, exceeding the largest \mbh\ ($10^{9.5}$\msun) measured for nearby galaxies. Shields et al. (2006a) derive a space density of 200 Gpc$^{-3}$ for black holes with \mbh\ $ > 10^{9.7} \msun$ based on broad line widths in the most luminous QSOs. Wyithe \& Loeb (2003) find a similar result based on the assumption that all QSOs are radiating at their Eddington limit. However, from the \mbhsigstar\ relationship (Gebhardt et al. 2000a; Ferrarese \& Merritt 2000; Tremaine et al. 2002), a black hole with mass $10^{9.7}$ \msun\ corresponds to a velocity dispersion of 500 \kms, larger than any \sigstar\ observed in local galaxies. Indeed, in a volume of space $\sim0.5$~Gpc$^3$ (approximately corresponding to the SDSS DR2; Strauss et al. 2002) none appear to have $\sigstar \gae 430$ \kms\ (Bernardi et al. 2006, 2008, hereafter B06 and B08 respectively).  These seemingly contradictory results raise the question of where the largest black holes reside, and what value of \sigstar\ characterize their host galaxies. 

Accurate knowledge of the largest values of \sigstar\ that occur in galaxies is therefore important for the understanding of the largest black holes in the universe, as well as having obvious bearing on the evolution of the largest galaxies. However, the signal-to-noise (S/N) of survey spectra---especially in the case of higher redshift galaxies---is often too low to make an accurate determination of \sigstar\ or to confidently rule out doubles and superposition as causes for large apparent velocity dispersions. With this goal in mind, we have selected eight galaxies which have amongst the highest velocity dispersions in the sample of B08 and appear, in \hst\ images and SDSS spectra, to be single galaxies. We have obtained high S/N spectra with the Hobby-Eberly Telescope (HET) to measure \sigstar\ with the greater accuracy and to double-check the likelihood of superposition.

Absolute magnitudes used in this study are calculated using the cosmological parameters $\hnot = 70~\kmspmpc, \Omega_{\rm M} = 0.3$, and $\Omega_{\Lambda} = 0.7$.

\section{Sample Selection}
\label{s:sample}

\begin{figure}
\begin{center}
\psfig{file=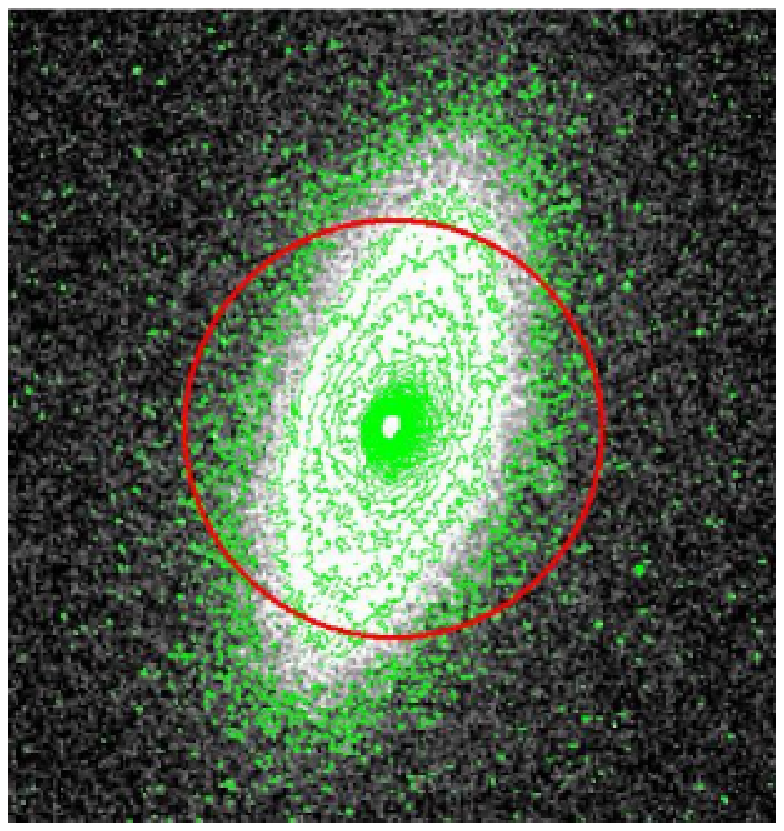,width=5cm}\psfig{file=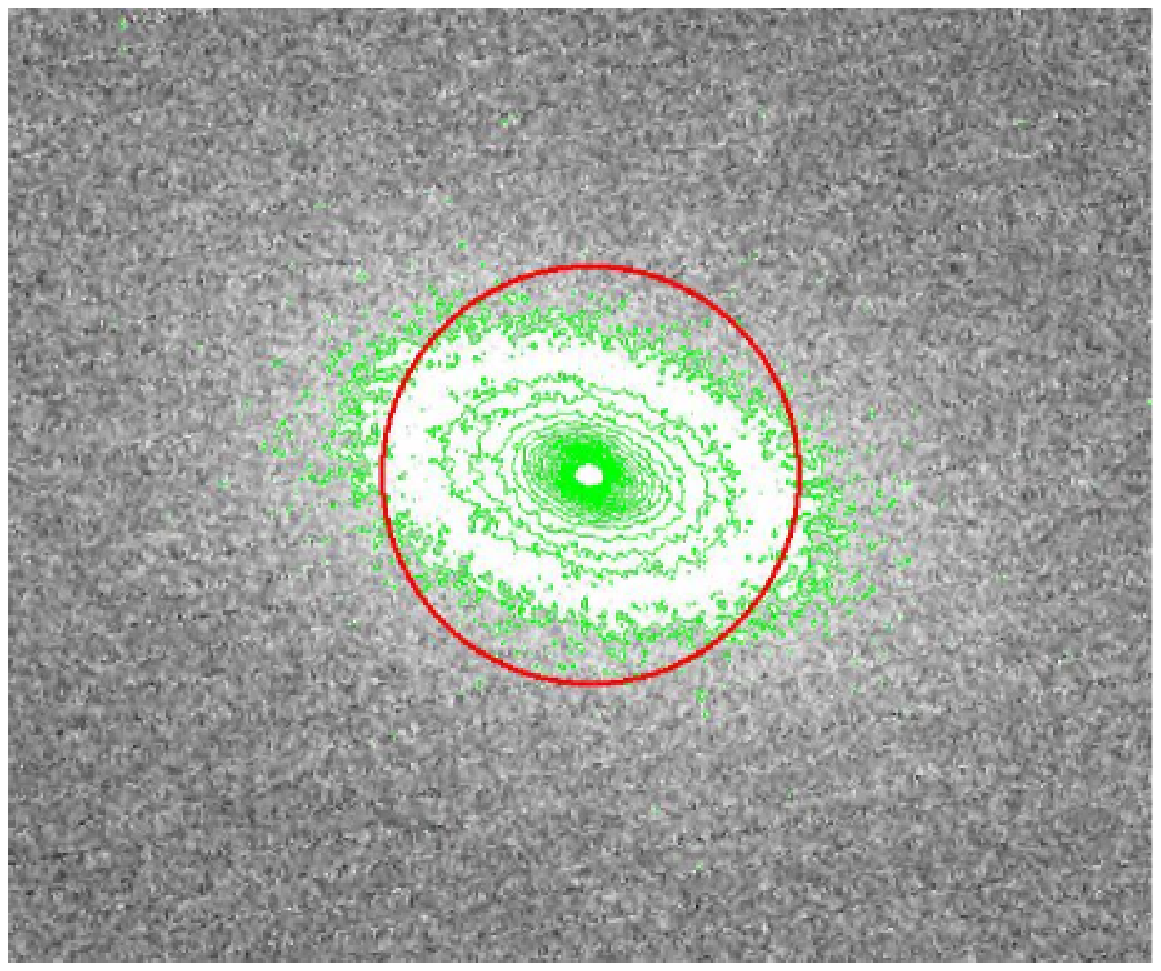,width=5cm}
\psfig{file=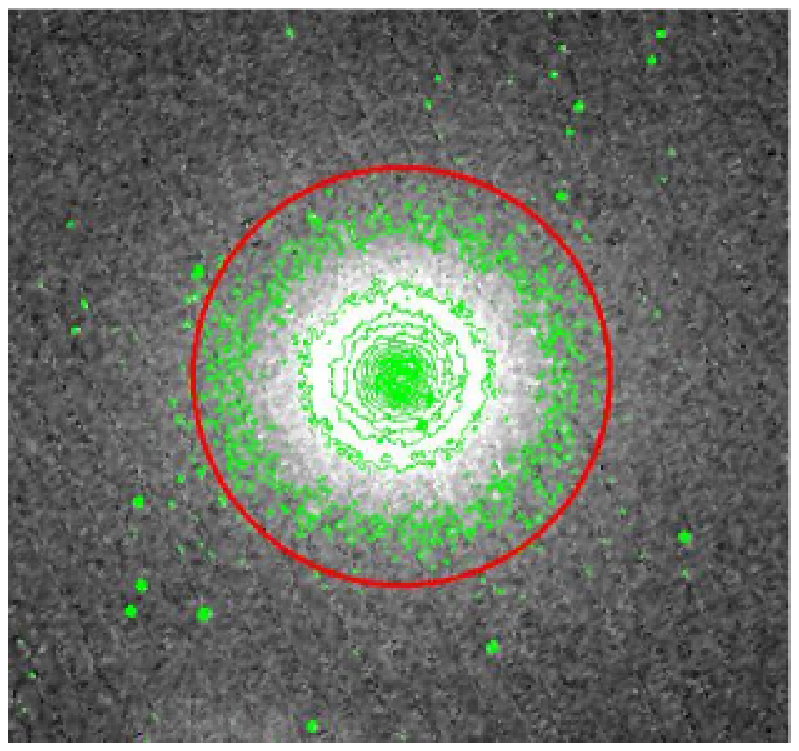,width=5cm}\psfig{file=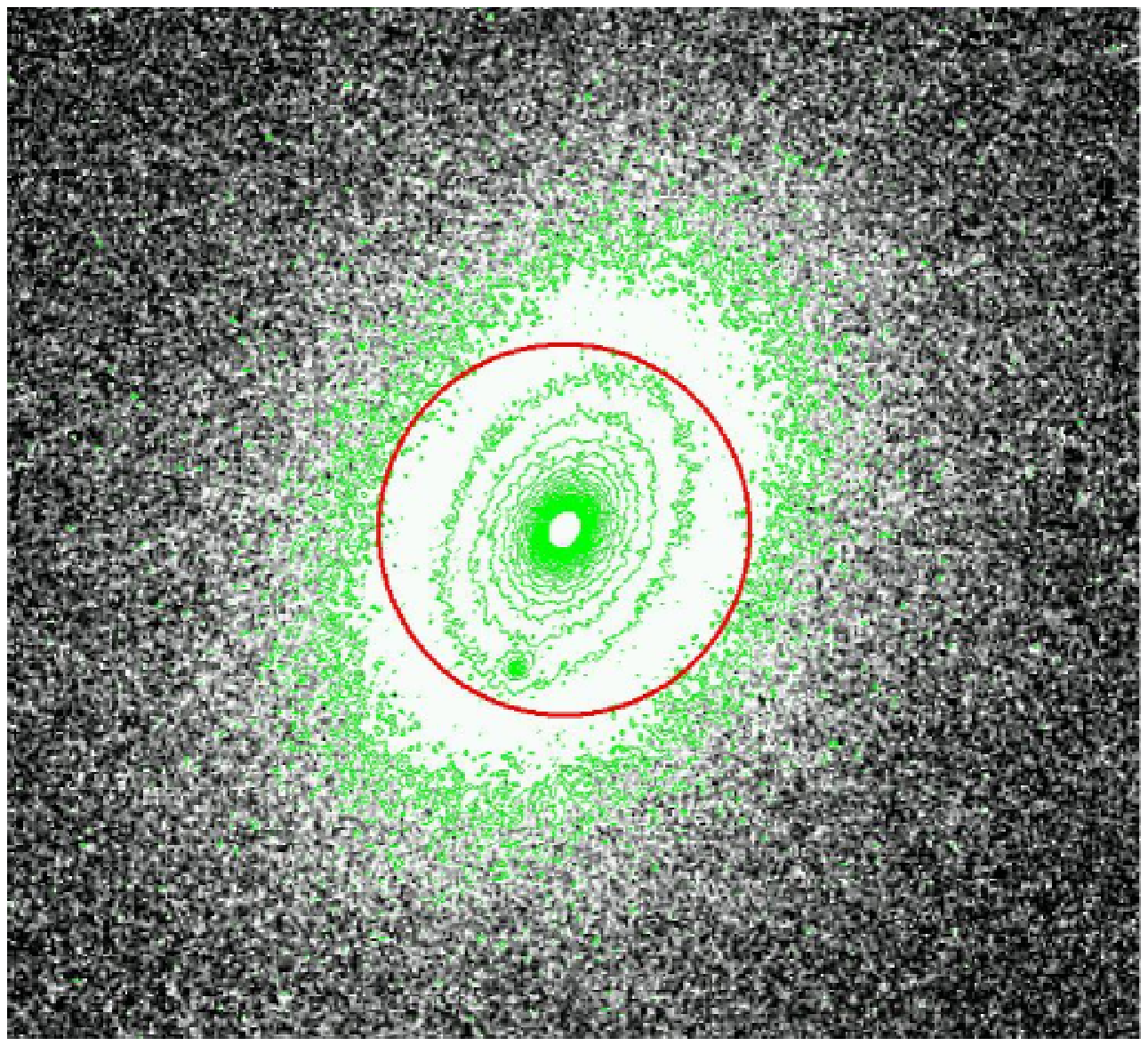,width=5cm}
\psfig{file=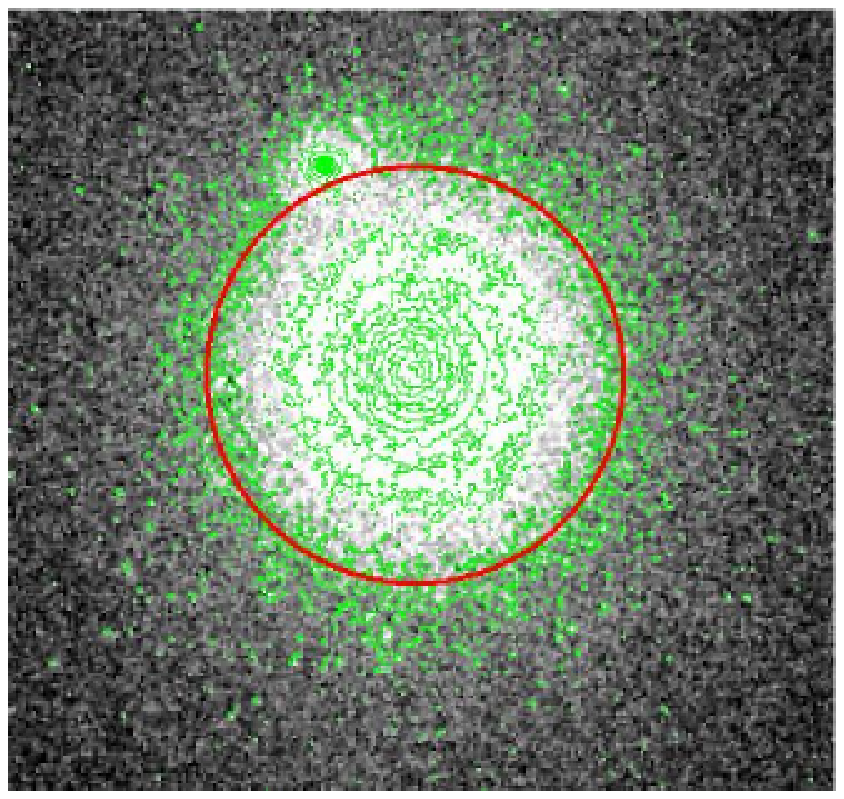,width=5cm}\psfig{file=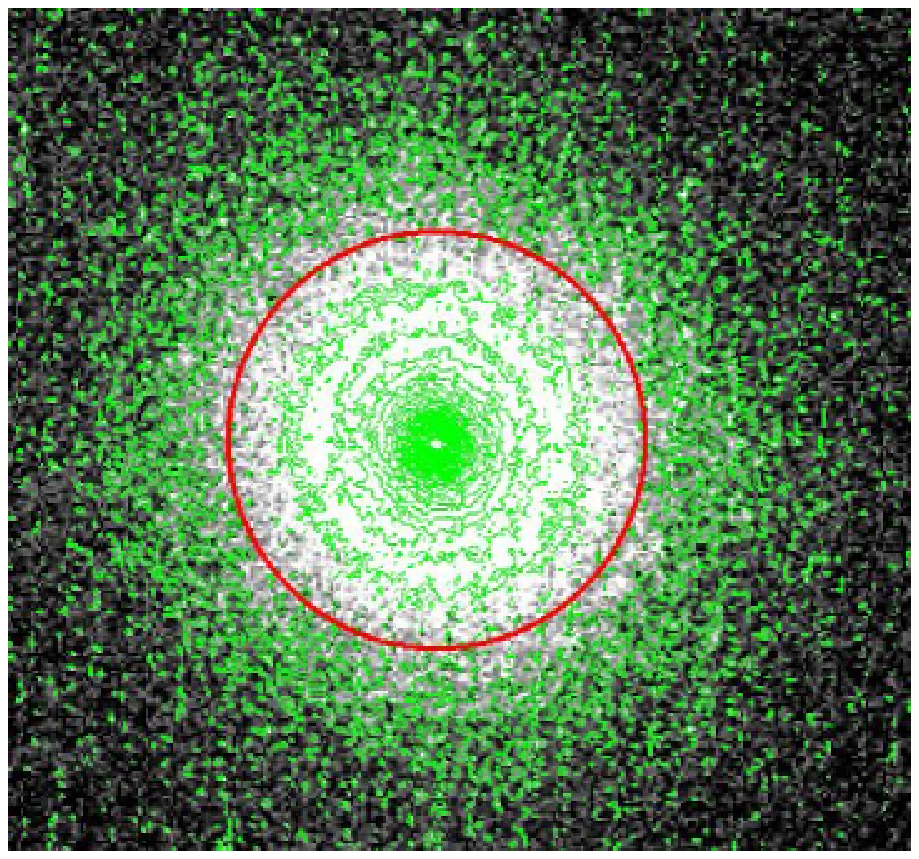,width=5cm}
\psfig{file=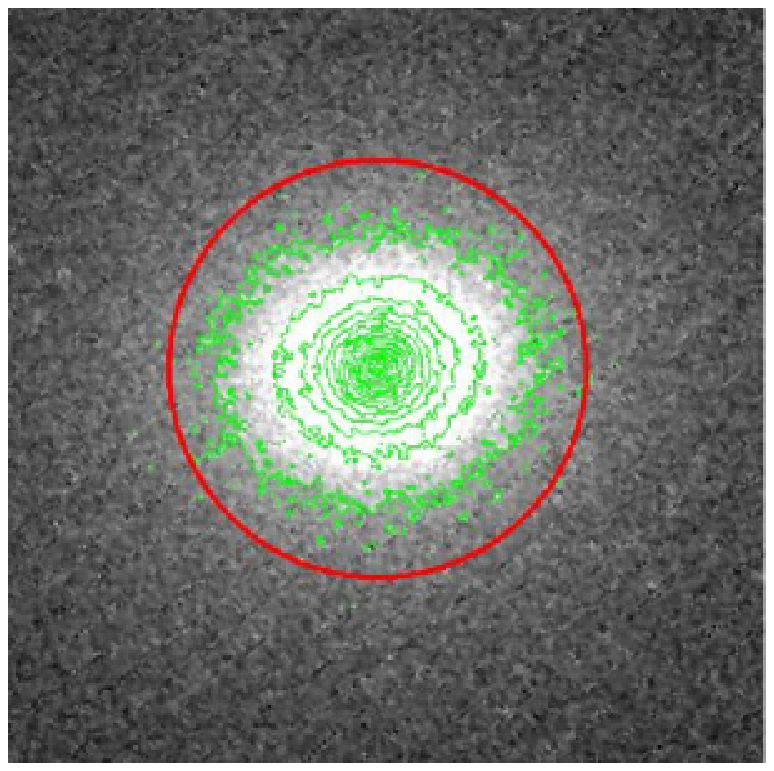,width=5cm}\psfig{file=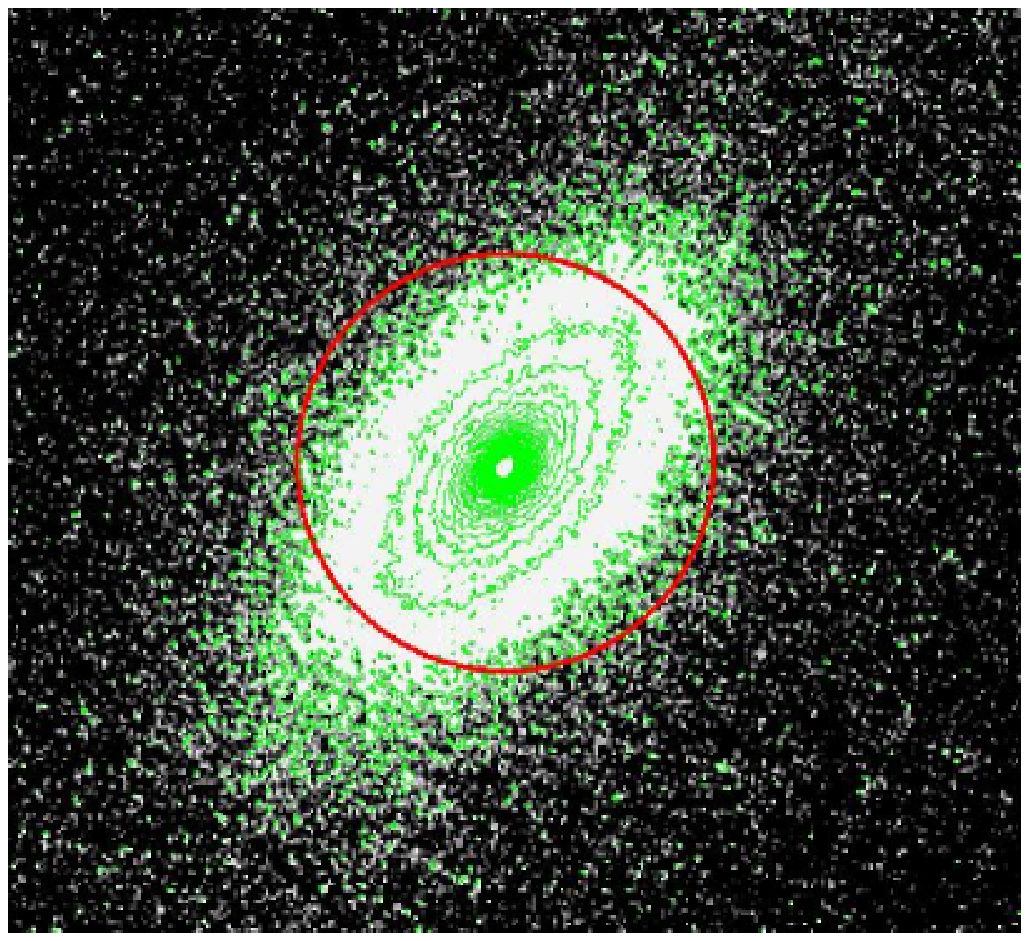,width=5cm}
\end{center}
\caption{\hst\ images from B08 for the galaxies in Table \ref{t:results}. Red circles show the size of the 3\arcs\ SDSS fiber.}
\label{f:images}
\end{figure}

B06 have studied high-\sigstar\ galaxies in a sample from the SDSS approximately corresponding to Data Release 2 with $z \leq 0.3$, corresponding to a co-moving volume of approximately 0.5 Gpc$^{-3}$. The selection criteria are described in detail in B06. Briefly, they chose early-type galaxies with Petrosian apparent magnitudes in the range $14.5 \leq r_{\mathrm Pet} \leq 17.75$ and measured velocity dispersions in the SDSS spectroscopic pipeline. Out of a total of $\sim40,000$ galaxies, they found 100 candidate galaxies with \sigstar\ $ > 350$ \kms. Of these, roughly half are superpositions as shown by the SDSS images or spectrum line profiles. For most of the remaining objects, B08 obtained \hst\ images that in turn reveal roughly half of them to be superpositions. The surviving 23 objects have values of \sigstar\ ranging up to $\sim 430$ \kms\ as measured from the SDSS spectra. We selected eight of these objects based on their large \sigstar\ and availability in the sky during the proposed observing period. Figure \ref{f:images} shows \hst\ images of these galaxies from B08.  

\section{Observations and Data Reduction}
\label{s:obs}

Observations were obtained using the Marcario Low Resolution Spectrograph (LRS; Hill et al. 1998) on the 9.2-m Hobby-Eberly Telescope during the period spanning April - November, 2006. We used a slit width of 2\arcs, a 600 line mm$^{-1}$ grating, and a GG385 blocking filter; the $3072 \times 1024$ Ford Aerospace CCD has a plate scale of 0.235 arcsec/pixel and was binned 2 $\times$ 2. Exposure times were 40 minutes each. Wavelength coverage for this setup is 4300 - 7300 \AA, with a resolving power $R = \lambda/\Delta\lambda = 650$. The instrumental width is $\sigma_{\mathrm{inst}} = $ FWHM/2.35 = 200 \kms. On most nights the sky conditions were spectroscopic with $1.5 - 2.0$ arcsecond seeing. Table \ref{t:galaxies} lists the spectral extraction apertures and galaxy morphologies. 

The data reductions were carried out with {\sc fortran} algorithms developed by one of the co-authors (K.G.) using standard reduction techniques. We first performed CCD corrections for overscan. Flat fields were created from five images of an internal continuum source and normalized, and each image was then divided by the appropriate flat field. Rectification was performed by determining the $x$ and $y$ locations of the spectra on the images and approximating a trace. Images were sky subtracted using apertures for the night sky that were defined manually for each object. Images for each galaxy were combined and then a 1-dimensional spectrum was extracted by defining an aperture for the object and summing pixel values for each wavelength bin. The wavelength scale was calibrated by means of a fourth-order polynomial wavelength solution for an emission spectrum of a Cd lamp.

\subsection{Measuring \sigstar}

\begin{figure}
\begin{center}
\psfig{file=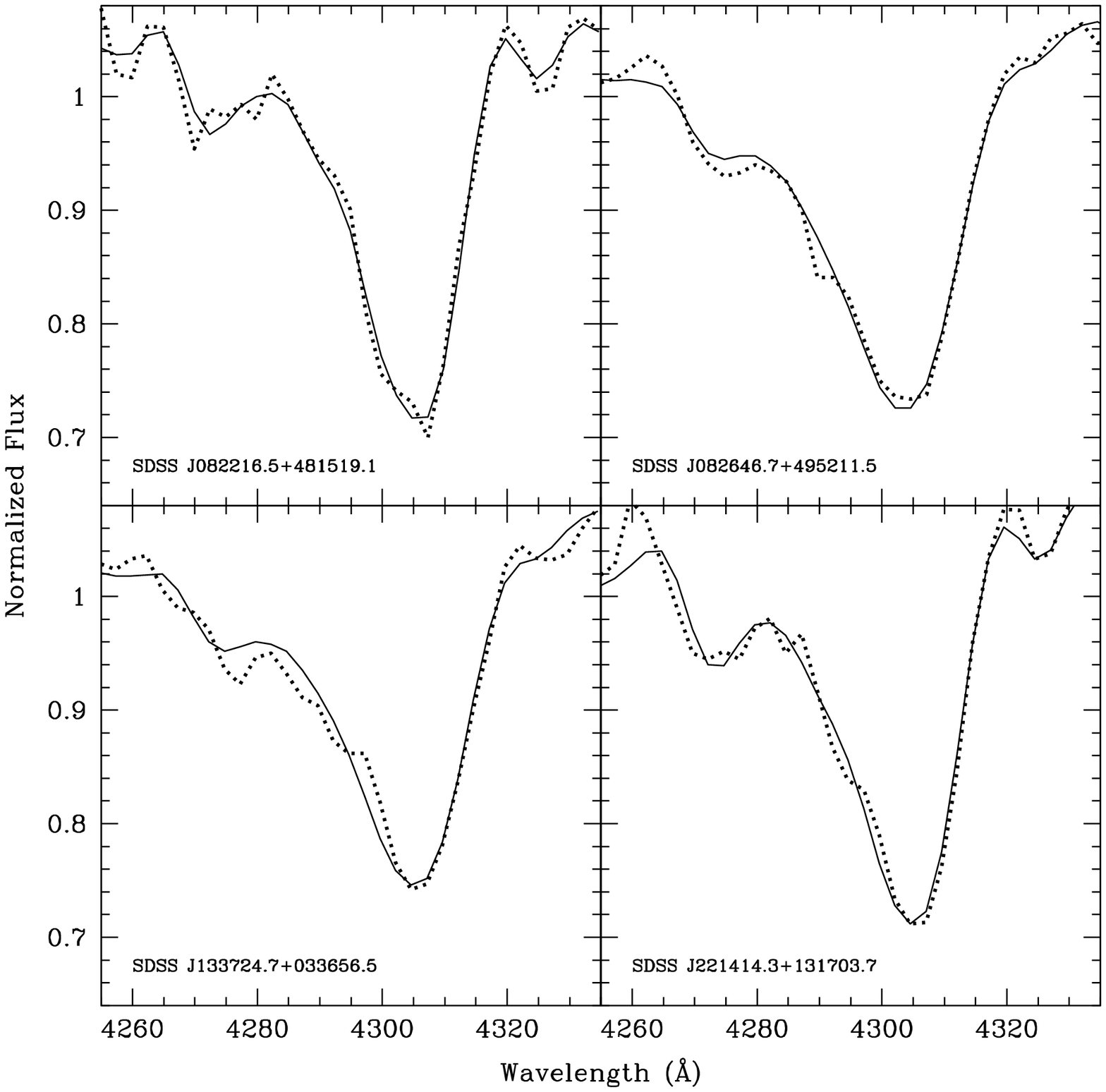,width=8cm}\psfig{file=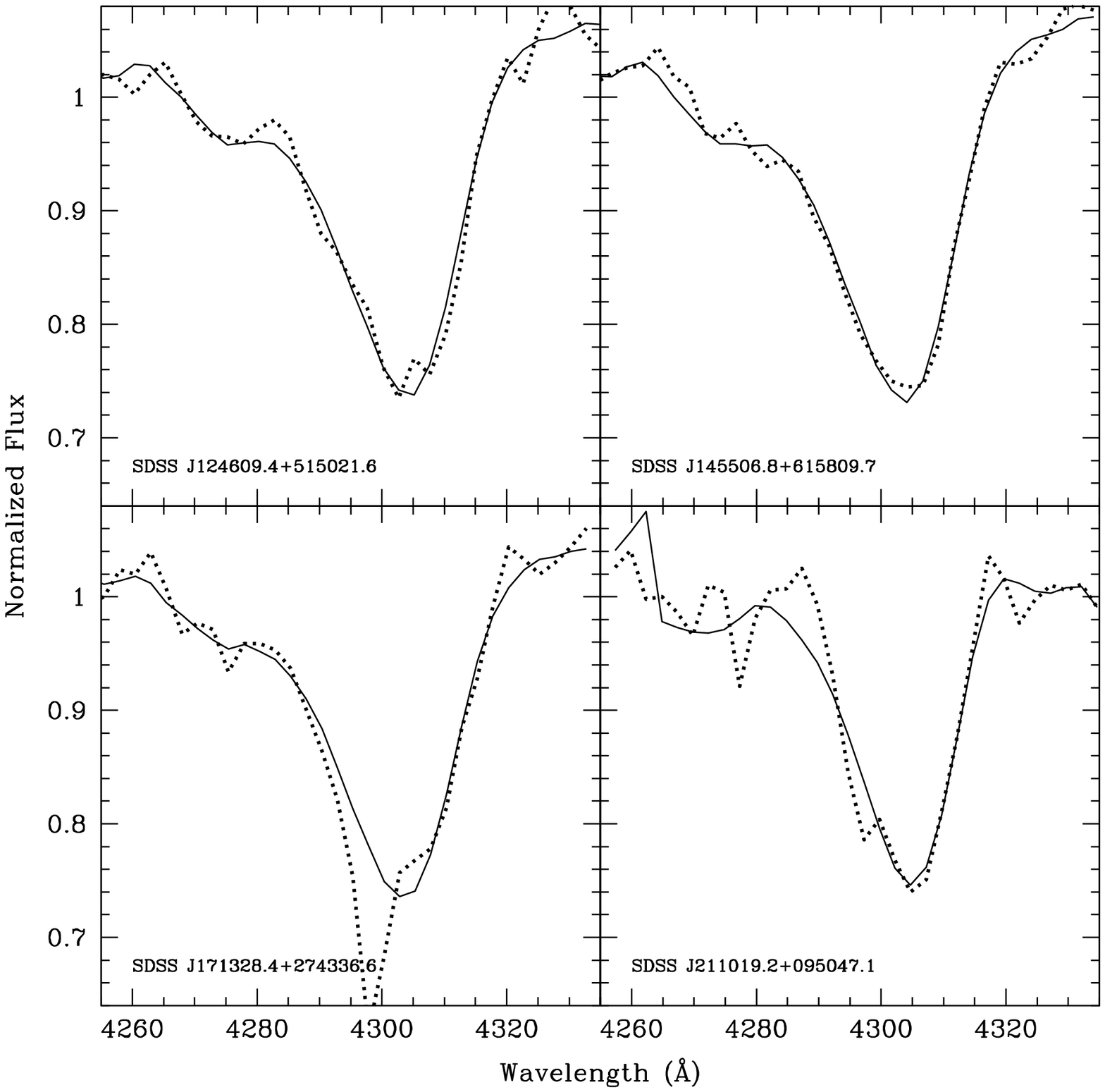,width=8cm}
\end{center}
\caption{Fits to the G-band region of the spectrum for the galaxies in Table \ref{t:results}. The left panel shows fits for power-law galaxies; the right panel shows fits for core galaxies.}
\label{f:spectra}
\end{figure}

Details of the spectrum modeling are described in Gebhardt et al. (2000b). Briefly, we used an automated procedure to fit the spectrum separately in the regions of the G-band line and the Ca II H+K lines using an adjustable combination of stellar templates to simulate the spectrum of an elliptical galaxy. Table \ref{t:stars} lists the template stars and their spectral types. There is some difficulty in simulating spectra for large-\sigstar\ galaxies, because the stellar population is not well known. For this reason, we constrained our fitting routine to the regions immediately surrounding the G-band and Ca II H+K lines. A measurement for \sigstar\ was obtained through a simultaneous fitting of the continuum and the velocity profiles of the G-band line or the Ca II H+K lines until an optimal match to the galaxy spectrum was achieved (see Figure \ref{f:spectra}). We modeled the velocity profiles using two methods: (1) least square fits of a pure Gauss function; and (2) least square fits of a Gauss-Hermite function, which is Gaussian multiplied by Hermite polynomials. The coefficients of the Hermite polynomials, $h_3$ and $h_4$, characterize the asymmetrical and symmetrical deviations from a pure Gaussian profile, respectively (see van der Marel and Franx 1993). The first two moments of the Gauss-Hermite represent the mean velocity and velocity dispersion, \sigstar. We quote the \sigstar\ measurements from fitting method (1), and used the parameters of the fit from fitting method (2) as a check for binarity or other irregularities. Goodness of fit was confirmed by visual inspection of the model fit and velocity profiles for each galaxy. 

Error bars were determined via Monte Carlo trials in which the initial data were randomly varied based on the rms noise per pixel and then re-fitted by our automated procedure to get a \sigstar. The standard deviation of \sigstar\ from 100 such trials determined the 1$\sigma$ level of error. The errors for the G-band fits and Ca II H+K fits are comparable, but we quote only the \sigstar\ of the G-band measurement, since it is less susceptible to template mismatch due to $\alpha$-element enhancement (see Barth et al. 2002). 

\section{Results}
\label{s:results}

\begin{figure}
\begin{center}
\psfig{file=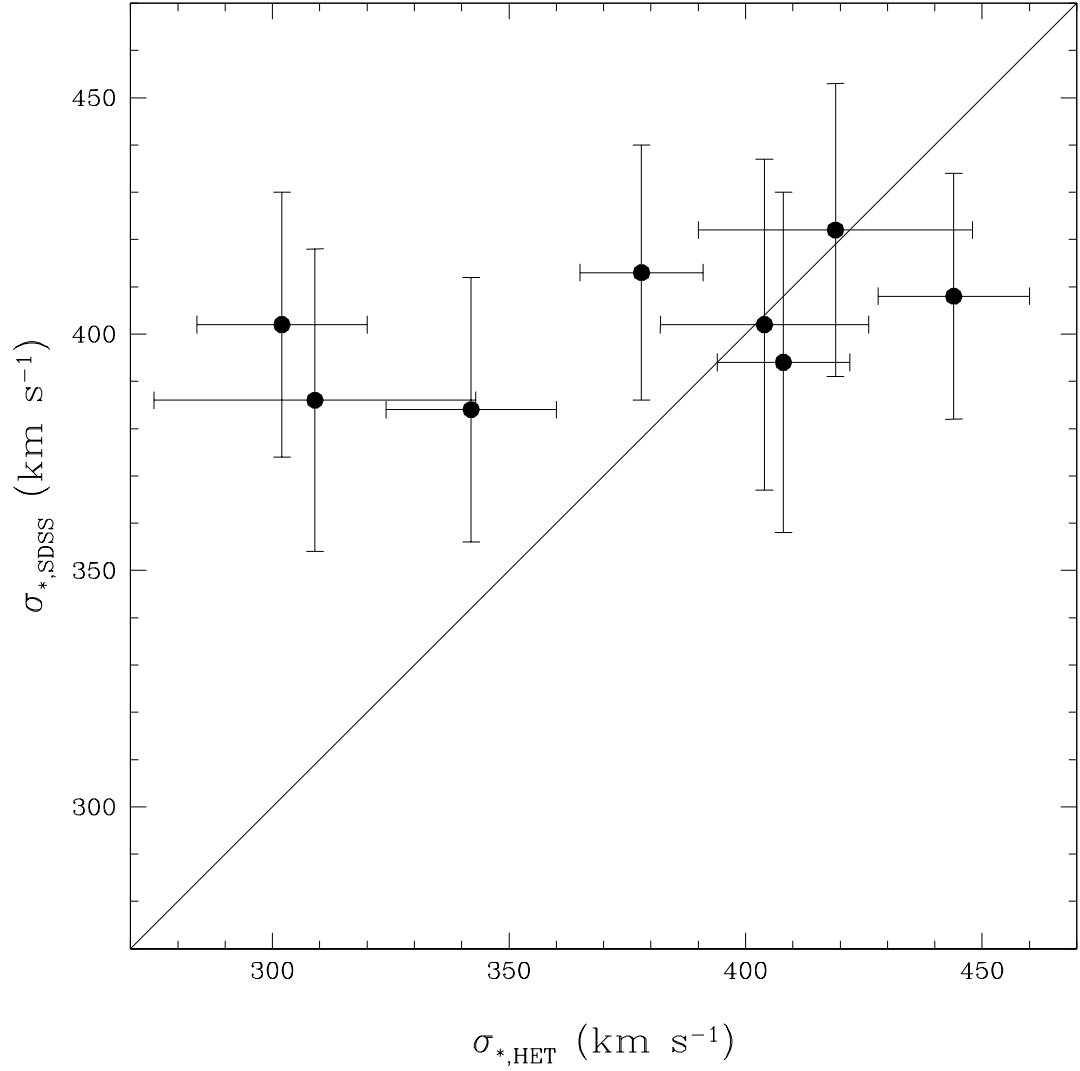,width=15cm}
\end{center}
\caption{$\sigstar(\mathrm{SDSS})$ versus $\sigstar(\mathrm{HET})$. The error bars show the 1$\sigma$ level of error. The solid line shows the 1:1 relationship.
\label{f:sigma}}
\end{figure}

Table \ref{t:results} lists \sigstar\ measured for each galaxy from both the HET and SDSS spectrum. Figure \ref{f:sigma} shows that in most cases there is agreement within $\pm 10\%$. The largest \sigstar\ we find is 444 $\pm 16$ \kms\ for SDSS J082646.72+495211.5, which differs from the SDSS spectrum measurement by $\sim 9\%$. The smallest \sigstar\ in our sample is 302 $\pm 18$ \kms\ for SDSS J082216.57+481519.1, which differs from the SDSS spectrum measurement by $\sim 25\%$. Figure \ref{f:sigma} shows that the largest discrepancies are for lower-\sigstar\ objects. We applied our fitting algorithm to the SDSS spectra, and our measured dispersions agree with those reported in B06. There is no obvious reason for the discrepancies. One possibility may be that the galaxies were selected for large \sigstar\ from SDSS. Given the steeply dropping number of galaxies with increasing \sigstar, and the large number of galaxies in the full SDSS dataset, noise may act to elevate a few objects from $\sigstar \approx 300$ \kms\ to a measured value of $\sigstar \gae 400$ \kms. 

%We selected galaxies with the largest dispersions in the SDSS; the range in \sigstar\ from the SDSS sample is small (384 - 422 \kms). Since there are many more galaxies with $\sigstar = 300$ \kms, it is likely that some galaxies with \sigstar\ much larger than this are due to noisy spectra. There are only a few galaxies in the SDSS with $\sigstar > 400$ \kms, so it is unlikely that noisy spectra would account for dispersions much larger than this. 

\subsection{Tests for Binarity}

We attempted to assess the likelihood of binarity by inspecting plots of the velocity profiles for asymmetries and multiple components---flat-topped profiles in particular indicate binaries. The strength of the Hermite coefficients $h_3$ and $h_4$ served as numerical indicators of binaries: non-zero $h_3$ indicates asymmetry, with $h_3 < 0$ corresponding to a blue wing on the profile and $h_3 > 0$ corresponding to a red wing; $h_4 < 0$ indicates a boxy or flat-topped profile. The $|h_3|$ and $|h_4|$ for each of the galaxies in our sample were modest ($< 0.1$), and none of the velocity profiles showed indication of asymmetry or multiple components. However, this method of assessing binarity appears to detect only obvious binaries with very large measured \sigstar. We fit the SDSS spectra of the galaxies in Table 2 of B06, which are flagged as superpositions, to see whether $h_3$ and $h_4$ would help to distinguish them as binaries. We found that only in cases where measured $\sigstar \gae 500$ \kms\ and visual inspection of the profile fits showed obvious signs of binarity did the $|h_3|$ and $|h_4|$ exceed 0.1. We rely on the \hst\ images and lack of gross effects on the line profiles to rule out binaries, but cannot rule out precisely superimposed images and modest velocity separations that could enhance measured \sigstar.  

\subsection{The \sigstar-$L$ Relationship}

\begin{figure}
\begin{center}
\psfig{file=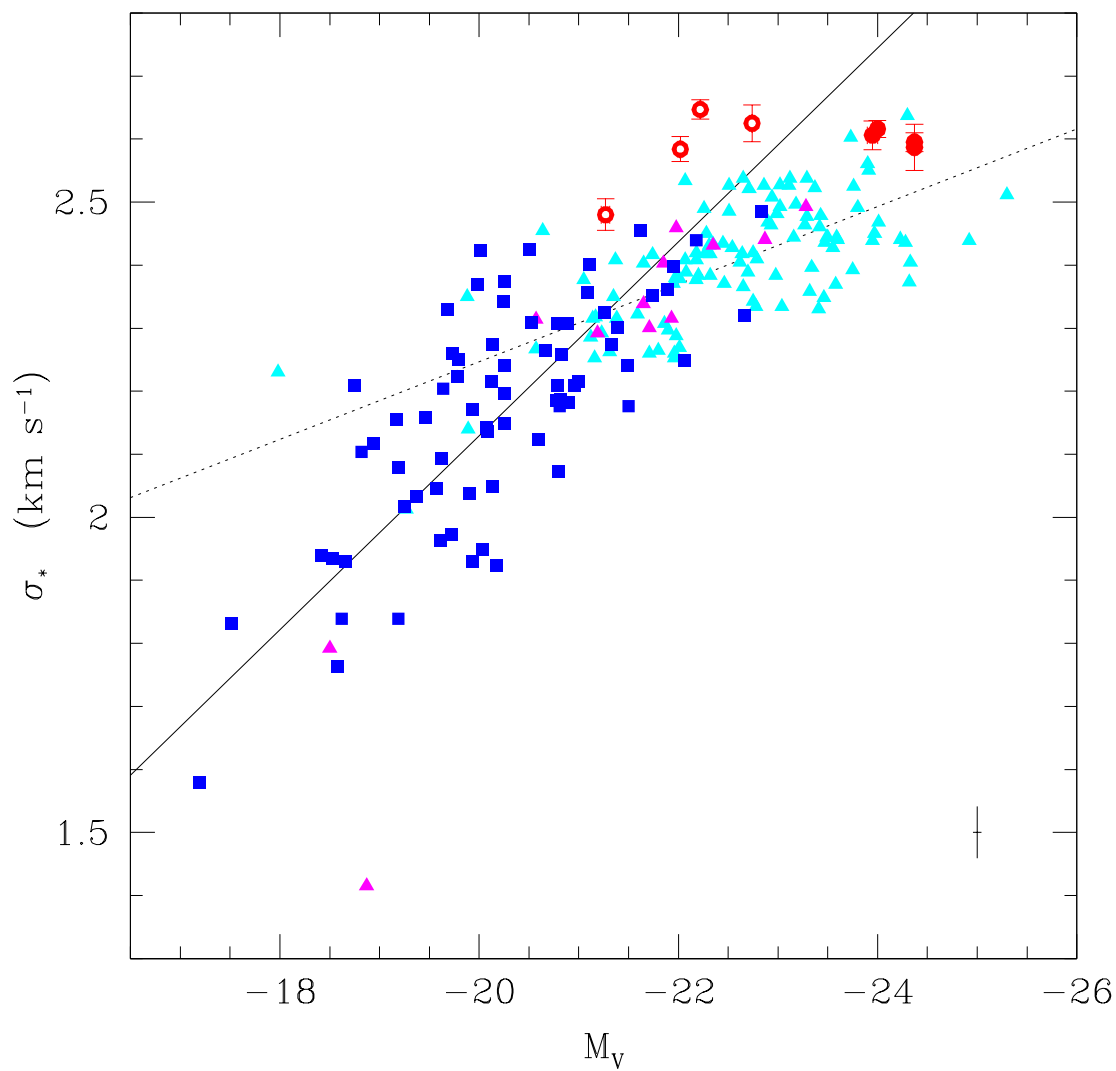,width=15cm,angle=0}
\end{center}
\caption{Measured \sigstar\ versus $V$-band absolute magnitude. Large circles show data for our galaxies. The leftmost four circles are power-law galaxies; the remaining circles indicate core galaxies. Error bars show the $1\sigma$ errors in \sigstar\ and $M_V$, however error bars for $M_V$ are smaller than the data points. The remaining data are from Figure~3 of Lauer et al. (2007a): squares show data for power-law galaxies; pentagons show data for intermediate type galaxies; triangles show data for core galaxies. The dashed line corresponds to $L \propto \sigma^7$ and the solid line to $L \propto \sigma^2$.}
\label{f:fj}
\end{figure}

In Figure \ref{f:fj}, we plot \sigstar\ against $V$-band absolute magnitude and compare our galaxies to the $\sigstar-L$ relationships and data from Figure 3 of Lauer et al. (2007a), which includes galaxies with core, power-law, and intermediate type profiles. Lauer et al. point out that the absolute magnitudes given for these galaxies by the SDSS catalog are underestimated due to improper sky subtraction in the SDSS pipeline. Hyde et al. (2008) corrected for this by carrying out their own photometric reductions of these objects, so we follow B08 in adopting their $r$-band absolute magnitudes, and convert to Johnson $V$-band magnitudes using the prescription from Jester et al. (2005) $(V = g - 0.59 [g - r] - 0.01)$. The four lowest luminosity galaxies agree with the $\sigstar-L$ relationship for power-law galaxies ($L \sim \sigstar^2$), while the higher luminosity galaxies agree with the relationship for core galaxies ($L \sim \sigstar^7$)---a direct analysis of the \hst\ light profiles confirms that the low $L$ objects are power-law galaxies, and the others are core galaxies (Hyde et al. 2008). Figure \ref{f:residual} shows the residual from the Faber-Jackson ($L \sim \sigstar^4$) relationship versus the residual from the effective radius--luminosity relationship (Bernardi et al. 2007a). More concentrated bulges are correlated with greater deviation from the Faber-Jackson relationship, in the sense that our galaxies have measured velocity dispersions larger than what is predicted from their luminosities. See Section \S \ref{s:disc} for the significance of this correlation.

\subsection{The \mbhsigstar\ Relationship}

Table \ref{t:results} also lists black hole masses predicted from the \mbhsigstar\ relationship for nearby galaxies quantified by Tremaine et al. (2002) and from the log-quadratic formulation of Wyithe et al. (2006). For our highest-\sigstar\ object, we predict \mbh\ $\approx 10^{9.5}$ \msun\ (Tremaine) and $\approx 10^{9.7}$ \msun\ (Wyithe), rivaling \mbh\ for M87 (Harms et al. 1994) and recently that of Abell 1836-BCG, the most massive black holes measured locally (Dalla Bont\`{a} et al. 2007). Note that the \sigstar\ of 375 \kms\ for M87 (Tremaine et al. 2002) places it above the \mbhsigstar\ relationship by a factor of about two in \mbh. If the \mbhsigstar\ relationship changes for high \mbh\ (e.g. Netzer et al. 2003), black hole masses could be larger for our galaxies than what is predicted here. Alternatively, \mbh\ may not correlate well with any galaxy properties at high \mbh. Given the scatter in the black hole - galaxy relationships, and lacking direct measurement of black hole masses $> 10^{9.5}$ \msun, the nature of the \mbhsigstar\ relationship remains uncertain for large \mbh.  

\section{Discussion}
\label{s:disc}

The largest \sigstar\ we measure is $\sim444$ \kms\ in a volume of space corresponding to 0.5~Gpc$^3$. We find no observational support for galaxies with $\sigstar \geq 500$ \kms, contrary to the QSO-based prediction of 100 such galaxies (see \S \ref{s:intro}). Evidently, either (1) black hole masses derived from QSOs are overestimated or (2) the linear relationship between log \mbh\ and log \sigstar\ breaks down for high \mbh\ (Netzer 2003). Regarding the first possibility, QSO black hole masses are typically estimated using the ``photoionization method'' (see Shields et al. 2003 and references therein), in which the virial mass is given by $\mbh = v^2 R / G$. The velocity is given by the width of broad line region (BLR) emission lines such as \hbeta, and the QSO luminosity provides an estimate of the radius of the BLR. Assuming the BLR gas is indeed virialized, the main uncertainties are the BLR radius-luminosity relationship and the geometry of the BLR (flattened or spherical). The $\rblr-L$ relationship was calibrated using the reverberation mapping technique on a sample of lower-luminosity AGN (see review by Kaspi 2007) and scales as $\rblr \sim L^{\gamma}$ with $\gamma = 0.5 - 0.7$. Bentz et al. (2006) find $\gamma = 0.52$ after correcting $L$ for host galaxy star light, consistent with photoionization physics (Shields et al. 2003). This relationship holds for a relatively wide range of luminosities up to $\lbol = 10^{46}$~\ergps, and is extrapolated to higher luminosities to estimate black hole masses for bright QSOs. As discussed by Salviander et al. (2007), the calibration used by Shields et al. (2006a) is consistent with observed \mbh\ and \sigstar\ in Seyfert galaxies (Onken et al. 2004). The calibration recommended by Bentz et al. (2006) gives \mbh\ 0.1 dex {\em larger} than that of Shields et al. (2006a). Shields et al. point out that a majority of the highest luminosity QSOs in the sample of Shields et al. (2003) have $\mbh > 10^{9.7} \msun$, which argues against scatter in the derivation of \mbh\ from AGN luminosity and broad line width as a major source of the large \mbh\ for this sample. Thus, an overestimation of \mbh\ in QSOs appears unlikely as a cause of the conflict between the abundance of large black holes in QSOs and the number of galaxies with sufficiently high \sigstar. 

We must therefore consider the possibility of a breakdown in the linear relationship between log \mbh\ and log \sigstar\ at high \mbh. This is consistent with the properties of the brightest cluster galaxies (BCGs). As shown in Figure \ref{f:fj} and discussed by Lauer et al. (2007a), the Faber-Jackson relationship levels off for the most luminous galaxies (also see Oegerle \& Hoessel 1991; Bernardi et al. 2007a). This suggests that \mbh\ may track galaxy luminosity rather than \sigstar\ for the largest galaxies. Lauer et al. (2007a) have augmented the galaxy luminosity function of Blanton et al. (2003) to include brightest cluster galaxies (BCGs) from the Postman \& Lauer (1995) sample. This luminosity function (Lauer et al 2007b, Figure 1) gives $10^{2.1}$ galaxies per Gpc$^3$ with $M_V < -23.9$. If current black hole samples are unbiased (see discussion in Bernardi et al. 2007b), then this luminosity corresponds to $\mbh = 10^{9.7}$ \msun\ (see Equation 6 of Lauer et al. 2007a). Within the uncertainties, this is an adequate number of BCGs to host the largest black holes observed in QSOs. Note that Tundo et al. (2007) show that scatter in the $\mbh - \lbulge$ relationship can bias estimates like this upwards by factors of several. A consequence is that the largest black holes will typically reside in proportionally modest-sized galaxies (see discussion in \S \ref{s:evol}). This may introduce scatter into the black hole - host galaxy relationships for the largest black holes. 

\begin{figure}
\begin{center}
\psfig{file=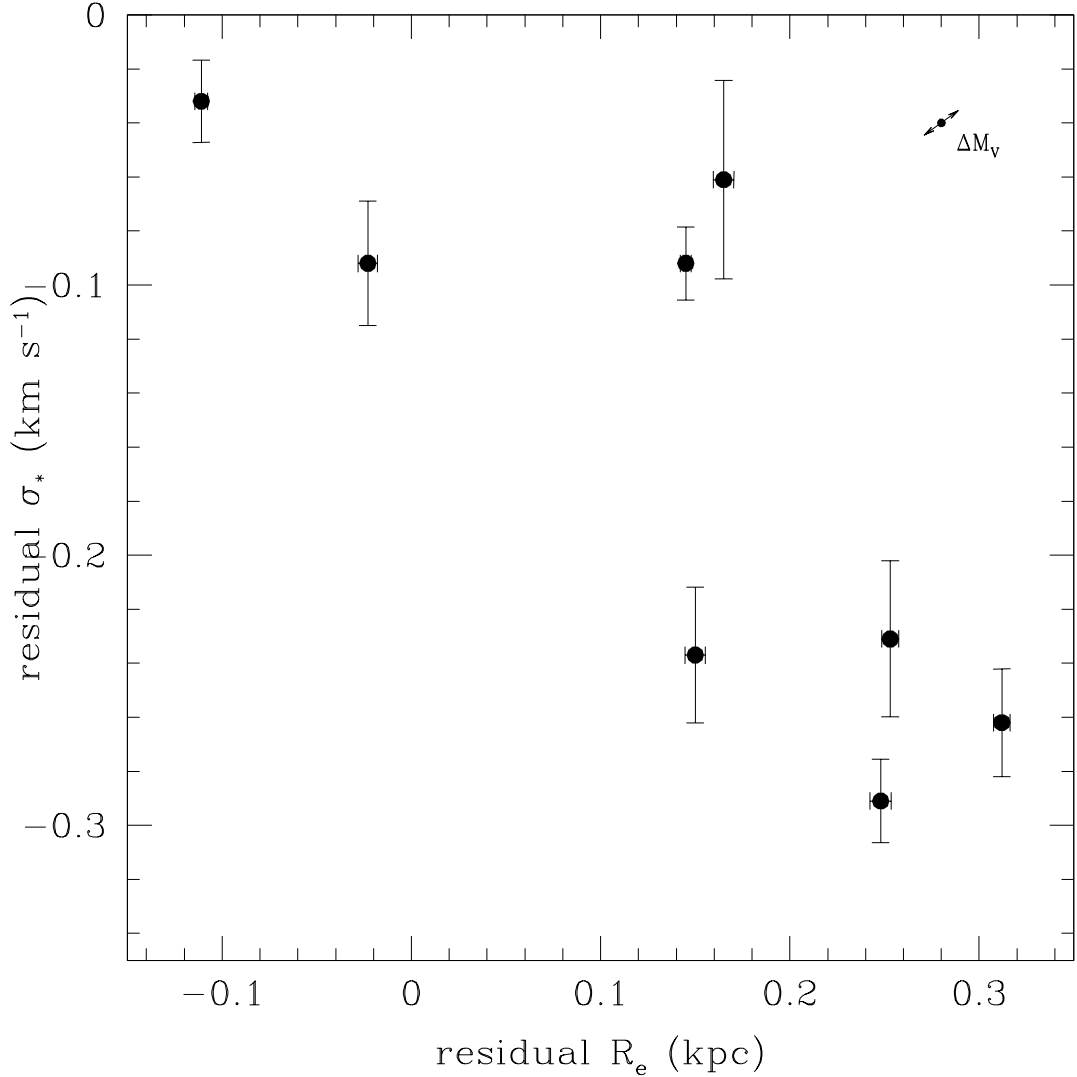,width=15cm,angle=0}
\end{center}
\caption[residual.eps]{Residuals from the bulge velocity dispersion--luminosity (Faber-Jackson) relationship versus residuals from the effective radius--luminosity relationship. Residual $\sigstar = $ log $\sigstar(L)$ $-$ log $\sigstar(\mathrm{HET})$ and residual $R_e = $ log $R_e(L)$ $-$ log $R_e$. Log $\sigstar(L)$ was calculated using the $\sigstar-L$ relationship (Equation 2 of Bernardi 2007) and log $R_e(L)$ was calculated using the $R_e-L$ relationship (Equation 2 of Bernardi et al. 2007a). The arrows in the upper right of the figure indicate the magnitude and sense of the change in the relationship that would be due to changes in the predicted $R_e$ and \sigstar\ should $M_V$ change by a factor equal to the average error in $M_V$, $\delta_{M_V} = 0.018$. 
\label{f:residual}}
\end{figure}

Does our highest velocity dispersion represent a physical limit for \sigstar? The $\sigstar-L$ relationship flattens out for \sigstar\ at high $L$, which may be explained by influences on \sigstar\ in galaxy merger models; the largest galaxies are believed to have formed through a succession of dissipationless, or ``dry,'' mergers. Loeb \& Peebles (2003) discuss possible causes of a limiting value of \sigstar. Major merger simulations by Boylan-Kolchin et al. (2006) show that the type of trajectory during galaxy mergers determines the concentration, and therefore \sigstar, of the remnant spheroid. For orbits with high angular momentum, more energy is transferred from the stellar component to the dark matter halo, resulting in a more concentrated remnant and a larger \sigstar. Figure \ref{f:residual} is consistent with this picture, showing more concentrated bulges (in the sense that the measured effective radius, $R_e$, is smaller than that predicted from the $R_e - L$ relationship) corresponding to over-large \sigstar\ (for a given bulge luminosity and the $L \sim \sigstar^4$ relationship). Since we use luminosity to predict both \sigstar\ and $R_e$, the
axes in Figure \ref{f:residual} are not independent. However, these correlated errors cannot explain the trend in the figure. Given the definition of the residuals, $\Delta$ log $\sigstar = $ log $\sigstar(L)$ $-$ log $\sigstar$ and $\Delta$ log $R_e = $ log $R_e(L)$ $-$ log $R_e$, and that $R_e \propto L^{0.5}$ and $\sigstar \propto L^{0.25}$, errors in luminosity will affect $\Delta$~log~$R_e$ and $\Delta$~log~$\sigstar$ in the sense of creating a positive slope in the figure rather than the negative slope that is actually seen. The magnitude and sense of the change in the relationship due to correlated errors is indicated in the figure. The negative slope in Figure \ref{f:residual} is qualitatively consistent with expectations from the virial theorem---for a given spheroid mass (and therefore luminosity) a decrease in radius corresponds to an increase in measured \sigstar. 

\section{Comment on Evolution of the \mbhsigstar\ Relationship}
\label{s:evol}

Recent studies by Shields et al. (2006b), Peng et al. (2006), Salviander et al. (2007), and others suggest that the \mbhsigstar\ relationship evolves with redshift in the sense that, for a given \mbh, the velocity dispersion is too small at higher redshifts. Whether this apparent evolution is real is controversial. For example, Salviander et al. (2007) studied \mbh\ and \sigstar\ in SDSS QSOs at redshifts up to $z = 1.2$, estimating \sigstar\ from the width of \oiii~$\lambda5007$ and \oii~$\lambda3727$. They found that higher redshift QSOs had \mbh\ too large for \sigstar\ by $\sim0.5$~dex. The interpretation of this trend requires caution. Brighter QSOs generally have larger black holes, so that larger \mbh\ corresponds in practice to higher redshift and larger look-back time. (1) Salviander et al. (2007) and Lauer et al. (2007b) describe how a Malmquist-like bias leads to over-representation of large black holes in modest galaxies. This bias is greater at the steep, high luminosity end of the QSO luminosity function; and it may give rise to the appearance of evolution in the \mbhsigstar\ relationship. (2) An observational bias favoring the detection of narrow lines affects the statistics of the \oiii\ and \oii\ line widths. Salviander et al. find that these two biases account for about 0.3~dex of their apparent evolution, leaving $\sim0.15$~dex of residual evolution. However, as discussed above, \sigstar\ levels off in the $\sigstar-L$ (Faber-Jackson) plot (Figure \ref{f:fj}) at high luminosity, where brightest cluster galaxies (BCGs) do not follow the relationship (Oegerle \& Hoessel 1991; Lauer et al. 2007a; Bernardi et al. 2007a). Use of the local \mbhsigstar\ relationship (Tremaine et al. 2002) with $\mbh \propto \sigstar^4$ at high \mbh\ could overestimate the expected \sigstar\ and give the appearance of evolution. A correction of only 0.1~dex on this basis applied to the residual evolution of Salviander et al. (2007) would leave no significant remaining evolution. Alternatively, Bernardi et al. (2007b) suggest that there is little bias in the \mbhsigstar\ relationship. Instead, they suggest that bias occurs in the $\mbh - \lbulge$ relationship, as a result of a steeper slope in the $\sigstar - L$ relationship for black hole samples compared to the SDSS sample. 
 
\section{Conclusion}
\label{s:conclusion}

We have used the HET LRS to obtain high signal-to-noise spectra for eight galaxies selected from the sample of B06--B08 representing galaxies with the highest measured values for \sigstar\ that appear to be single galaxies in \hst\ images and SDSS spectra. The maximum \sigstar\ we find is 444~km/s, similar to B06. We find no support for \sigstar\ $\ge 500$ \kms. If black hole masses are reliable, this is indicative of a \mbhsigstar\ relationship that deviates at high \mbh\ from the locally-observed log-linear relationship. Though our data do not address this issue directly, it is consistent with the change in the relationship between \sigstar\ and luminosity for the brightest elliptical galaxies.  

\acknowledgements

We thank John Kormendy and Milo$\check{\mathrm{s}}$ Milosavljevi\'{c} for helpful discussions. G.A.S. gratefully acknowledges the hospitality of Lick Observatory and the support of the Jane and Roland Blumberg Centennial Professorship in Astronomy. M.B. acknowledges support from NASA LTSA-NNG06GC19G grant. We thank the SDSS team for the enormous effort involved in conducting the survey and making the results conveniently accessible to the public.

Funding for the Sloan Digital Sky Survey (SDSS) has been provided by the Alfred P. Sloan Foundation, the Participating Institutions, the National Aeronautics and Space Administration, the National Science Foundation, the U.S. Department of Energy, the Japanese Monbukagakusho, and the Max Planck Society. The SDSS Web site is http://www.sdss.org/. The SDSS is managed by the Astrophysical Research Consortium (ARC) for the Participating Institutions. The Participating Institutions are The University of Chicago, Fermilab, the Institute for Advanced Study, the Japan Participation Group, The Johns Hopkins University, the Korean Scientist Group, Los Alamos National Laboratory, the Max-Planck-Institute for Astronomy (MPIA), the Max-Planck-Institute for Astrophysics (MPA), New Mexico State University, University of Pittsburgh, University of Portsmouth, Princeton University, the United States Naval Observatory, and the University of Washington.

\newpage

\begin{deluxetable}{lc}
\tablewidth{0pt}
\tablecaption{Template Stars\label{t:stars}}
\tablehead{
\colhead{Star}  &
\colhead{Spectral Type}}
\startdata
HD10761 & G8 III\\
HD111812 & G0 III\\
HD161797 & G5 IV\\
HD199960 & G1 V\\
HD219449 & K0 III\\
HD220954 & K1 III\\
HD39283 & A2 V\\
HD81146 & K2 III\\
HD85235 & A3 IV\\
HD92588 & K1 IV\\
G166-45 & A5 V\\
\enddata

\tablecomments{Stars contributing to the stellar template used by the fitting algorithm to simulate the continuum of an elliptical galaxy.}
\end{deluxetable}

\newpage

%\begin{rotate}
\begin{deluxetable}{lccc}
\tabletypesize{\scriptsize} 
\tablewidth{0pt}
\tablecaption{Spectral extraction apertures and galaxy profile types \label{t:galaxies}}
\tablehead{
\colhead{Galaxy}  &
\colhead{aperture} &
\colhead{aperture} &
\colhead{profile} \\
\colhead{(SDSS J)} &
\colhead{(arcseconds)} & 
\colhead{(kpc)} & 
\colhead{ }}  
\startdata
082216.5$+$481519.1 & 2.35 & 5.34 & power-law \\
082646.7$+$495211.5 & 2.82 & 7.74 & power-law \\
124609.4$+$515021.6 & 1.88 & 7.76 & core \\
133724.7$+$033656.5 & 2.35 & 5.55 & power-law \\ 
145506.8$+$615809.7 & 2.82 & 11.8 & core \\
171328.4$+$274336.6 & 2.12 & 9.38 & core \\
211019.2$+$095047.1 & 1.65 & 6.08 & core \\
221414.3$+$131703.7 & 2.82 & 7.51 & power-law \\
\enddata

\tablecomments{The HET spectra were extracted from the apertures shown in columns 2 and 3. Galaxy profile types are from Bernardi et al. (2008).} 

\end{deluxetable}

\newpage

%\begin{rotate}
\begin{deluxetable}{lccccccccccc}
\tabletypesize{\scriptsize} 
\tablewidth{0pt}
\tablecaption{Physical parameters for galaxies\label{t:results}}
\tablehead{
\colhead{Galaxy}  &
\colhead{$z$}  &
\colhead{$M_V$}  &
%\colhead{$M_r$}  &
\colhead{$\sigma_*$} & 
\colhead{$\sigma_*$} &
\colhead{log \mbh}  &
\colhead{log \mbh} &
\colhead{log \mbh} &
\colhead{log $R_e$} \\
\colhead{(SDSS J)}  &
\colhead{ }  &
\colhead{ }  &
\colhead{(HET, G-band)}  &
\colhead{(SDSS)}  &
\colhead{(\sigstar)} &
\colhead{(\sigstar, quad.)} &
\colhead{($L$)} &
\colhead{ } \\
\colhead{ }  &
\colhead{ }  &
\colhead{(mag)}  &
%\colhead{(mag)}  &
\colhead{(\kms)}  &
\colhead{(\kms)}  &
\colhead{($M_{\odot}$)} &
\colhead{($M_{\odot}$)} &
\colhead{($M_{\odot}$)} &
\colhead{(kpc)}}
\startdata
082216.5$+$481519.1 & 0.127 & -21.27 $\pm$ 0.042 & 302 $\pm$ 18 & 402 $\pm$ 28 & 8.85 & 8.84 & 8.28 & 0.337 \\
082646.7$+$495211.5 & 0.159 & -22.22 $\pm$ 0.009 & 444 $\pm$ 16 & 408 $\pm$ 26 & 9.52 & 9.69 & 8.79 & 0.547 \\
124609.4$+$515021.6 & 0.269 & -23.95 $\pm$ 0.052 & 404 $\pm$ 22 & 402 $\pm$ 35 & 9.36 & 9.47 & 9.70 & 1.249 \\
133724.7$+$033656.5 & 0.133 & -22.74 $\pm$ 0.004 & 419 $\pm$ 29 & 422 $\pm$ 31 & 9.43 & 9.57 & 9.06 & 0.645 \\
145506.8$+$615809.7 & 0.274 & -24.37 $\pm$ 0.015 & 408 $\pm$ 14 & 394 $\pm$ 36 & 9.31 & 9.42 & 9.92 & 1.460 \\
171328.4$+$274336.6 & 0.297 & -24.00 $\pm$ 0.010 & 378 $\pm$ 13 & 413 $\pm$ 27 & 9.40 & 9.52 & 9.73 & 1.108 \\
211019.2$+$095047.1 & 0.230 & -24.37 $\pm$ 0.004 & 309 $\pm$ 34 & 386 $\pm$ 32 & 9.28 & 9.37 & 9.92 & 1.093 \\
221414.3$+$131703.7 & 0.153 & -22.02 $\pm$ 0.008 & 342 $\pm$ 18 & 384 $\pm$ 28 & 9.27 & 9.36 & 8.68 & 0.391 \\
%
%082216.5$+$481519.1 & 0.127 & -21.53 & -21.27 $\pm$ 0.042 & 302 $\pm$ 18 & 402 $\pm$ 28 & 8.85 & 8.84 & 8.28 & 0.337 \\
%082646.7$+$495211.5 & 0.159 & -22.66 & -22.22 $\pm$ 0.009 & 444 $\pm$ 16 & 408 $\pm$ 26 & 9.52 & 9.69 & 8.79 & 0.547 \\
%124609.4$+$515021.6 & 0.269 & -24.24 & -23.95 $\pm$ 0.052 & 404 $\pm$ 22 & 402 $\pm$ 35 & 9.36 & 9.47 & 9.70 & 1.249 \\
%133724.7$+$033656.5 & 0.133 & -23.04 & -22.74 $\pm$ 0.004 & 419 $\pm$ 29 & 422 $\pm$ 31 & 9.43 & 9.57 & 9.06 & 0.645 \\
%145506.8$+$615809.7 & 0.274 & -24.73 & -24.37 $\pm$ 0.015 & 408 $\pm$ 14 & 394 $\pm$ 36 & 9.31 & 9.42 & 9.92 & 1.460 \\
%171328.4$+$274336.6 & 0.297 & -24.34 & -24.00 $\pm$ 0.010 & 378 $\pm$ 13 & 413 $\pm$ 27 & 9.40 & 9.52 & 9.73 & 1.108 \\
%211019.2$+$095047.1 & 0.230 & -24.36 & -24.37 $\pm$ 0.004 & 309 $\pm$ 34 & 386 $\pm$ 32 & 9.28 & 9.37 & 9.92 & 1.093 \\
%221414.3$+$131703.7 & 0.153 & -22.32 & -22.02 $\pm$ 0.008 & 342 $\pm$ 18 & 384 $\pm$ 28 & 9.27 & 9.36 & 8.68 & 0.391 \\
%
\enddata

\tablecomments{$M_V$ calculated using $r$- and $g$-band magnitudes from Hyde et al. (2008) and the transformation from Jester et al. (2005). Errors for $M_V$ were calculated by adding in quadrature weighted rms errors for $M_r$ and $M_g$. \sigstar$_{\mathrm{,SDSS}}$ source: Bernardi et al. (2006). \mbh(\sigstar) and \mbh(\sigstar-quad.) inferred from the \mbhsigstar\ relationship quantified by Tremaine et al. (2002) and Wyithe (2006), respectively; \mbh($L$) inferred from the $\mbh-L$ relationship quantified by Lauer et al. (2007a). $R_e$ is from Hyde et al. (2008).}

\end{deluxetable}
%\end{rotate}

\newpage

% hst images of our galaxies

\newpage

% spectra for our galaxies

\newpage

% sigma(SDSS) vs sigma(HET)

\newpage

% our objects on F-J (from Lauer)

\newpage

% F-J residuals vs L-Re residuals

\end{document}